\documentclass[11pt,leqno]{article}
\usepackage{multicol}

\usepackage{psfig}
\usepackage{xspace}

\usepackage{amsmath}

\makeatletter%
\def\nottoobig#1{{\hbox{$\left#1\vcenter to1.111\ht\strutbox{}\right.\n@space$}}}
\makeatother%

\makeatother%

\makeatletter%

\newcount\hour  \newcount\minutes  \hour=\time  \divide\hour by 60
\minutes=\hour  \multiply\minutes by -60  \advance\minutes by \time
\def\mmmddyyyy{\ifcase\month\or Jan\or Feb\or Mar\or Apr\or May\or Jun\or Jul\or
  Aug\or Sep\or Oct\or Nov\or Dec\fi \space\number\day, \number\year}
\def\hhmm{\ifnum\hour<10 0\fi\number\hour :%
  \ifnum\minutes<10 0\fi\number\minutes}
\def\Draft{{\it Draft of \mmmddyyyy}}

\def\ps@jtsheadings{%
\def\@oddhead{\it\rightmark\hfil\rm\thepage}%
\def\@oddfoot{\hfil\Draft}%
\if@twoside%
\def\@evenhead{\rm\thepage\hfil\it\leftmark}%
\def\@evenfoot{\Draft\hfil}%
\else
\let\@evenhead\@oddhead%
\let\@evenfoot\@oddfoot%
\fi%
}
\def\ps@jtsplain{%
\def\@oddhead{\hfil\Draft}%
\def\@oddfoot{\hfil\rm\thepage\hfil}%
\let\@evenfoot\@oddfoot%
\if@twoside \def\@evenhead{\Draft\hfil} \else \let\@evenhead\@oddhead \fi
}

\def\chaptermark#1{\markboth{\thechapter.\ #1}{\thechapter.\ #1}}%
\def\sectionmark#1{\markright{\thesection.\ #1}}

\def\section{\@startsection {section}{1}{\z@}
    {3.5ex plus1ex minus.2ex}{2.3ex plus.2ex}{\Large\bf}}
\def\subsection{\@startsection{subsection}{2}{\z@}
    {3.25ex plus1ex minus.2ex}{1.5ex plus.2ex}{\large\bf}}
\def\subsubsection{\@startsection{subsubsection}{3}{\z@}
    {3.25ex plus1ex minus.2ex}{1.5ex plus.2ex}{\normalsize\bf}}
\def\paragraph{\@startsection{paragraph}{4}{\z@}
    {3.25ex plus1ex minus.2ex}{1em}{\normalsize\bf}}
\def\subparagraph{\@startsection{subparagraph}{4}{\parindent}
    {3.25ex plus1ex minus.2ex}{1em}{\normalsize\bf}}

\makeatother%

\makeatletter \@beginparpenalty=10000 \makeatother

\def\underl#1 {\leavevmode\let\first=\relax\underli #1 }
\def\underli#1 {\ifx&#1\let\next=\relax\unskip
                \else\let\next=\underli\first\ulinebox{#1}\fi\let\first=\undersp\next}
\def\undersp{\penalty50\ulinebox{\space}\penalty50}
\def\ulinebox#1{\vtop{\hbox{\strut#1}\hrule}}%
\def\unice#1 {\underl #1 & }
\def\desclabel#1{\bf #1\hfil}
\def\desc{\list{}{%
\labelwidth= \leftmargin
\advance \labelwidth by -\labelsep
\let \makelabel=\desclabel}}



\makeatletter %

\newlength{\filength}
\newsavebox{\gcbox}
\sbox{\gcbox}{\framebox[\filength]{\rule{0ex}{2ex}}}

\newlength{\leftjustindent}
\newlength{\@leftjustindent}
\def\leftjust{\let\\\@leftjustcr\let\end\@endleftjust
  \addtolength{\@leftjustindent}{\leftjustindent} \vcenter\bgroup
\halign\bgroup \hbox to\displaywidth{
\rule{\@leftjustindent}{0ex}$\displaystyle##$\hfill }\crcr }
\def\endleftjust{\crcr\egroup\egroup\endgroup}
\def\@endleftjust#1{\crcr\egroup\egroup\@checkend{#1}\endgroup}
\def\@leftjustcr{\crcr}

\newcommand{\singlespacing}{\let\CS=
\@currsize\renewcommand{\baselinestretch}{1}\tiny\CS}
\newcommand{\singlespacingplus}{\let\CS=
\@currsize\renewcommand{\baselinestretch}{1.25}\tiny\CS}
\newcommand{\doublespacing}{\let\CS=
\@currsize\renewcommand{\baselinestretch}{1.75}\tiny\CS}
\newcommand{\draftspacing}{\let\CS=
\@currsize\renewcommand{\baselinestretch}{2.0}\tiny\CS}

\makeatother%

\singlespacingplus

\mathcode`\0="0030      %
\mathcode`\1="0031
\mathcode`\2="0032
\mathcode`\3="0033
\mathcode`\4="0034
\mathcode`\5="0035
\mathcode`\6="0036
\mathcode`\7="0037
\mathcode`\8="0038
\mathcode`\9="0039
\makeatletter
\clubpenalty=\@highpenalty
\widowpenalty=\@highpenalty
\makeatother

\makeatletter
\newcommand{\niceonespacing}{\let\CS=\@currsize\renewcommand{\baselinestretch}{1.1}\tiny\CS}\newcommand{\nicetwospacing}{\let\CS=\@currsize\renewcommand{\baselinestretch}{1.2}\tiny\CS}
\newcommand{\nicethreespacing}{\let\CS=\@currsize\renewcommand{\baselinestretch}{1.3}\tiny\CS}
\newcommand{\singlespacingplusplus}{\let\CS=\@currsize\renewcommand{\baselinestretch}{1.35}\tiny\CS}
\newcommand{\nicefourspacing}{\let\CS=\@currsize\renewcommand{\baselinestretch}{1.4}\tiny\CS}
\newcommand{\nicefivespacing}{\let\CS=\@currsize\renewcommand{\baselinestretch}{1.5}\tiny\CS}
\newcommand{\nicesixspacing}{\let\CS=\@currsize\renewcommand{\baselinestretch}{1.6}\tiny\CS}
\makeatother

\makeatletter
\def\@cite#1#2{[#1\if@tempswa , #2\fi]}
\makeatother

\makeatletter
\def\@citex[#1]#2{\if@filesw\immediate\write\@auxout{\string\citation{#2}}\fi
  \def\@citea{}\@cite{\@for\@citeb:=#2\do
    {\@citea\def\@citea{,\linebreak[0]}\@ifundefined
       {b@\@citeb}{{\bf ?}\@warning
       {Citation `\@citeb' on page \thepage \space undefined}}%
\hbox{\csname b@\@citeb\endcsname}}}{#1}}
\makeatother

\makeatletter
\def\ps@thesis{\def\@oddhead{\hfil\rm\thepage\hfil}\def\@oddfoot{}\def\@evenhead{\hfil\rm\thepage\hfil}\def\@evenfoot{}\def\chaptermark##1{}\def\sectionmark##1{}}
\makeatother

\makeatletter
\def\foobarpt{\textfont\z@\tenrm 
  \scriptfont\z@\ninrm \scriptscriptfont\z@\sevrm
\textfont\@ne\tenmi \scriptfont\@ne\ninmi \scriptscriptfont\@ne\sevmi
\textfont\tw@\tensy \scriptfont\tw@\ninsy \scriptscriptfont\tw@\sevsy
\textfont\thr@@\tenex \scriptfont\thr@@\tenex \scriptscriptfont\thr@@\tenex
\def\unboldmath{\everymath{}\everydisplay{}\@nomath\unboldmath
          \textfont\@ne\tenmi 
          \textfont\tw@\tensy \textfont\lyfam\tenly
          \@boldfalse}\@boldfalse
\def\boldmath{\@ifundefined{tenmib}{\global\font\tenmib\@mbi\@magscale1\global
        \font\tensyb\@mbsy \@magscale1\global\font
         \tenlyb\@lasyb\@magscale1\relax\@addfontinfo\@xiipt
              {\def\boldmath{\everymath
                {\mit}\everydisplay{\mit}\@prtct\@nomathbold
                \textfont\@ne\tenmib \textfont\tw@\tensyb 
                \textfont\lyfam\tenlyb\@prtct\@boldtrue}}}{}\@xiipt\boldmath}%
\def\prm{\fam\z@\tenrm}%
\def\pit{\fam\itfam\tenit}\textfont\itfam\tenit \scriptfont\itfam\ninit
   \scriptscriptfont\itfam\sevit
\def\psl{\fam\slfam\tensl}\textfont\slfam\tensl 
     \scriptfont\slfam\tensl \scriptscriptfont\slfam\tensl
\def\pbf{\fam\bffam\tenbf}\textfont\bffam\tenbf 
   \scriptfont\bffam\ninbf \scriptscriptfont\bffam\ninbf 
\def\ptt{\fam\ttfam\tentt}\textfont\ttfam\tentt
   \scriptfont\ttfam\nintt \scriptscriptfont\ttfam\nintt 
\def\psf{\fam\sffam\tensf}\textfont\sffam\tensf
    \scriptfont\sffam\tensf \scriptscriptfont\sffam\tensf
\def\psc{\@getfont\psc\scfam\@xiipt{\@mcsc\@magscale1}}%
\def\ly{\fam\lyfam\tenly}\textfont\lyfam\tenly 
   \scriptfont\lyfam\ninly \scriptscriptfont\lyfam\sevly
 \@setstrut \rm}

\makeatother

\def\land{{\; \wedge \;}}

\newenvironment{block}{\begin{list}{\hbox{}}{\leftmargin 1em
    \itemindent -1em \topsep 0pt \itemsep 0pt \partopsep 0pt}}{\end{list}}

\title{On Bounded-Weight Error-Correcting Codes\protect\thanks{Supported
in part by NSF grants
9322513,
9513368/\protect\linebreak[0]DAAD-315-PRO-fo-ab,
9701911,
and
9815095/DAAD-315-PPP-g\"u-ab.
Email:
{\protect\tt{}\{rbent,mschear,lane,istrate\}@}\protect\linebreak[0]\mbox{\protect\tt{}cs.rochester.edu}.}}
\author{ {\em  Russell Bent},
\ {\em  Michael Schear},
\ {\em  Lane A. Hemaspaandra},
\ and {\em  Gabriel Istrate}
\\
Department of Computer Science\\University of Rochester\\
             Rochester, NY 14627
}

\date{June 1, 1999\protect \\ ACM Subject Codes: H.1.1, E.4}

\makeatletter
\def\@listI{\leftmargin\leftmargini \parsep 4.5pt plus 1pt minus 1pt\topsep
6pt plus 2pt minus 2pt \itemsep  2pt plus 2pt minus 1pt}

\let\@listi\@listI
\@listi
\makeatother

\makeatletter %
 \newcommand{\setoffdisplay}{\rule{5.9in}{1pt}}

\makeatother



\begin{document}

\typeout{WARNING:  BADNESS used to supress reporting.  Beware.}
\hbadness=3000%
\vbadness=10000 %

\bibliographystyle{alpha}

\pagestyle{empty}
{\singlespacing\maketitle}
\begin{center}
{\large\bf Abstract}
\end{center}
\begin{quotation}
{\singlespacing

This paper computationally obtains optimal bounded-weight,
binary, error-correcting codes for a variety of 
distance bounds and dimensions.  
We compare the 
sizes of our codes to the sizes of optimal
constant-weight, binary, error-correcting codes, and 
evaluate the differences.

}
\end{quotation}

\pagestyle{plain}

\sloppy

\section{Introduction}

One goal of coding theory is to construct classes of codes
having optimal size. Studies have investigated versions of this problem for
classes of codes with various regularity properties, such as
linear codes over finite fields~\cite{binary:linear:codes}, binary 
self-dual codes~\cite{binary:self:dual:codes}, mixed binary-ternary 
codes~\cite{brouwer:codes}, and various classes of spherical
codes~\cite{spherical:codes:tables}. 

Two such important cases concern determining the values of $A(n,d)$ and
$A(n,d,w)$, where $A(n,d)$ is the 
number of codewords in the largest binary code of length $n$ having
minimum distance $d$, and $A(n,d,w)$ is the 
number of codewords in the largest binary code of
length $n$, minimum distance $d$, and weight $w$.
Optimal values for $A(n,d)$ and $A(n,d,w)$ have been tabulated by
Litsyn, Rains, and Sloane~\cite{binary:codes:table} and Sloane and 
Rains~\cite{constant:weight:codes:table}, respectively. 

It is conceivable that significant improvements in optimal
code size could be obtained by relaxing  
the restriction on the code weight in the definition of $A(n,d,w)$ 
from ``{\em equal} to $w$'' to ``{\em upper-bounded} by $w$,'' because
there would then be a greater number of words
potentially available for inclusion in the codes.
We present optimal, bounded-weight, binary, error-correcting codes for
a variety of  distance bounds and dimensions.  The method we employ
to obtain the optimal codes is based on the
observation that finding optimal bounded-weight codes can be
transformed to finding the size of a maximum clique in a suitably defined
graph. The clique-finding is accomplished primarily
using the branch and bound search
used in Brouwer et al.~(\cite{brouwer:codes},
see also~\cite{dimacs:graph:solvers} and the 
discussion later in this paper).

\section{Preliminaries}

Let $F$ be some finite set of characters---the {\em alphabet}.
A word of length $n$ over $F$ is an element of $F^n$. A code over $F$
of size $n$ is a set of words of length $n$ over $F$. 
A code over the alphabet $\{0,1\}$ is called {\em binary}.
Throughout this paper, we use the alphabet
$F=\{0,1\}$.

The distance, $d$, of a code is the smallest Hamming
distance between any two
codewords in the code. If we have two codewords, $x$ and $y$, both
of length $n$, we can represent these two words as $x_1 x_2 x_3
\cdots x_n$ and $y_1 y_2 y_3 \cdots y_n$, where 
$x_j$ is the $j^{th}$ bit in $x$. The Hamming distance between $x$ and
$y$ is the size of the set, $\left\{j:\; 1\leq j \leq n \land
x_{j} \neq y_{j}\right\}$.
The weight, $w$, of a binary word, $x$, is  equal to the
number of 1s in $x$. For a constant-weight~($w$) code, every word in the
code has the same weight, $w$.  In a bounded-weight ($w$) 
code, every
word has at most $w$ ones.

The standard 
reduction of finding optimal values of $A(n,d)$ and $A(n,d,w)$ to 
the problem of determining a maximum clique in a graph is 
as follows. The graph's vertices represent binary strings of
length $n$ (and legal weight, when appropriate).
Two vertices are joined by an edge if and only if their Hamming 
distance is at least $d$. 

It is easily seen that the connection between optimal code size and
maximum clique in a suitably constructed graph carries over to the
case of bounded-weight codes, and we indeed use exactly that in 
this paper.

\section{Results and Discussion}

The constant-weight bounds, many tight, tabulated by Sloane
were obtained from a variety of sources and 
methods~\cite{constant:weight:codes:table}.  An elegant
method for finding optimal codes of constant-weight is to use an
algebraic formula.   Methods of creating such formulas for 
certain cases are presented
in Brouwer et al.~\cite{brouwer:table}.  
No such algebraic formulas for instances of
bounded-weight codes are available yet.  In the absence of such a
method we tried various other methods
for obtaining good sets of codewords.  Many of the algorithms used were
bounded-weight variants of those suggested in the literature for
calculating good constant-weight
codes.  These methods included simulated 
annealing~\cite{gam-hem-shp-wei:j:annealing}, genetic 
algorithms~\cite{vaessens:gen-alg}, and a randomized greedy heuristic search.  
The codes generated by these methods were beaten or equaled by our final method
of obtaining codes, which was creating an
appropriate graph and seeking a large (in fact, usually maximum-size) clique
via different clique-finding algorithms.

Since the problem of finding a maximum clique in a graph has been
thoroughly investigated \cite{dimacs:clique}, it is natural to use a
reduction to this
problem as our basis for
finding good bounded-weight codes. The reduction is accomplished
by creating the graph
of possible codewords acceptable under the parameters for
length and weight. Each possible codeword is represented by a vertex
in
the graph. If two codewords have a proper Hamming distance, then an
edge is placed between them. The largest clique in the graph is
representative of a
maximum set of codewords such that the set meets all the
parameters.

We used two clique-finding algorithms suggested 
in Brouwer et al.~\cite{brouwer:codes}.
The first algorithm is a basic branch and bound search.  In the worst case,
it will search all possible combinations of nodes for cliques, but in
practice it keeps track of a best solution and travels only those paths
that have the potential to beat the current best solution.  This 
algorithm will always find a
maximum-size clique.
We used a publicly available coding
by Applegate and Johnson~(\cite{dimacs:graph:solvers}, 
see also~\cite{car-par:j:exact-cliques}).
The second algorithm
is a variant of 
semi-exhaustive greedy search.  This algorithm may not always find the
largest clique.  The algorithm begins by creating two sets of
nodes. The first set is nodes that are part of the clique
being created
and the second set is nodes that can
be added to the clique set without disrupting the
clique property of the set.
This available node set initially contains all the nodes
and the clique set is initially empty.  A node is chosen from
the nodes in the available set.  Those nodes that are not connected to
the chosen node are eliminated from the available set.  This process is
repeated until the number of nodes in the available set drops below a
user-defined threshold, $y$.  Once $y$ is reached, the
branch and bound algorithm is employed on the available set.  The
nodes are selected as follows. 
For a user-defined number $x$, $x$
nodes are chosen at random from the available node set.  The node with
the most edges in the set of $x$ nodes is
chosen.
We used a publicly available coding,
originally by Johnson, as modified by 
Applegate and Johnson~(\cite{dimacs:graph:solvers}, 
see also~\cite{ara-joh-mcg-sch:j:cliques-etc}).
For our
purposes, good 
results were achieved when $x = 0.1s$, where $s$ is the number
of nodes in the original graph, and $y = 100$.   We ran the algorithm
a thousand times in order to increase the odds of finding the largest clique.

The branch and bound algorithm was used on parameters
where the optimal constant-weight code sizes 
were known and the search spaces were small enough to allow results to
be obtained in reasonable amounts of time.  
For
example, it took forty-one CPU minutes to calculate $A(9,4,4)$
and this was considered reasonable.
On the other hand, the calculation of $A(9,4,7)$
was terminated
as it was taking an unreasonable amount of time.  
However,
running the greedy algorithm one thousand times on $A(9,4,7)$ took
just under seventy two CPU
minutes.\footnote{These
CPU times were obtained using a Sun Ultra 10.}

From our results, it is now clear 
that, with regards to changing from constant-weight to 
bounded-weight, there is little or no increase in number of codewords
in the best code until constant-weight codes become handicapped
with a decrease in search space. (As the weight of a constant-weight
code increases, the search space increases initially, but
then begins to decrease once $w > \lceil \frac{n}{2} \rceil$.
However, in the case of bounded-weight codes, the search space
continues to increase as $w$ approaches $n$.)
It is important to note that where there are
increases in the number of words in bounded-weight codes over
constant-weight codes, these new bounded-weight codes
can often be obtained trivially.  For example, if $w \ge d$, a
bounded-weight code can be
created by taking the constant-weight code at $A(n,d,w), w \ge d$ and
adding the 
word of all 0s.  This is because the word of all 0s has a Hamming
distance at least $d$ from all the words in the constant-weight code
$A(n,d,w)$, when $w \ge d$.
Other bounded-weight codes can be created in this manner by patching together
known constant-weight codes.

Clearly, a lower bound for bounded-weight codes is
$$\max_{m:0 \leq m < d} 
\left(\sum_{j : 0 \leq j \leq w \land (j
\equiv m~ (mod\; d))} A(n,d,j)\right).$$  
Results from the two clique-finding algorithms seem to usually merely 
meet this bound, and occasionally (see discussion below) beat it.
Tables~\ref{tab:dist4}, \ref{tab:dist6}, 
and~\ref{tab:dist8} illustrate these results.
It must be noted that the performance of the semi-exhaustive search
has only been tested on those parameters where the entire graph can be
created and stored in memory.  It remains to be seen if patched
codes can be matched or beaten easily in other cases.

\begin{table}[!t]
\begin{center}
\begin{tiny}
\begin{tabular}{|c|c||c||c|}
\hline
Length & Weight& Constant & Bounded \\
\hline
6 & 3 & 4 & 4\\
\hline
6 & 4 & 3  &4\\
\hline
6 & 5  & 1  & 4\\
\hline
6 & 6  & 1  & 4\\
\hline
\hline
7 & 3  & 7 & 7\\
\hline
7 & 4  & 7  & 8\\
\hline
7 & 5  & 3  & 8\\
\hline
7 & 6  & 1  & 8\\
\hline
7 & 7  & 1  & 8\\
\hline
\hline
8 & 3  & 8 & 8\\
\hline
8 & 4  & 14 & 15\\
\hline
8 & 5  & 8  & 15\\
\hline
8 & 6  & 4  & 16\\
\hline
8 & 7  & 1  & 16\\
\hline
8 & 8  & 1  & 16\\
\hline
\hline
9 & 3  & 12  & 12\\
\hline
9 & 4  & 18  & 19\\
\hline
9 & 5  & 18  &19$^{\star}$\\
\hline
9 & 6  & 12  &19$^{\star}$\\
\hline
9 & 7  & 4  &19$^{\star}$\\
\hline
9 & 8  & 1  &20$^{\star}$\\
\hline
9 & 9  & 1  &20$^{\star}$\\
\hline
\hline
10 & 3  & 13  & 13\\
\hline
10 & 4  & 30  & 31$^{\star}$\\
\hline
\hline
11 & 6 & 66  & 71$^{\star}$\\
\hline
\end{tabular}
\end{tiny}
\caption{\label{tab:dist4}Distance 4.
Note: The values superscripted with ``$\star$'' were
obtained through greedy search.}
\end{center}

\end{table}

\begin{table}[!t]
\begin{center}
\begin{tiny}
\begin{tabular}{|c|c||c||c|}
\hline
Length & Weight& Constant  & Bounded \\
\hline
8 & 4 &  2  &2\\
\hline
8 & 5  & 2  &2\\
\hline
8 & 6  & 1 & 2\\
\hline
8 & 7  & 1 & 2\\
\hline
8 & 8  & 1  & 2\\   
\hline
\hline
9 & 4  &  3  & 3\\
\hline
9 & 5  & 3  & 4\\
\hline
9 & 6  & 3   & 4\\
\hline
9 & 7  & 1  & 4\\
\hline
9 & 8  & 1  & 4\\
\hline
9 & 9  & 1  & 4\\
\hline
\hline
10 & 4 & 5  &5\\
\hline
10 & 5 & 6  &6\\
\hline
10 & 6 & 5  &6\\
\hline
10 & 7 & 3  & 6\\
\hline
10 & 8 & 1  & 6\\
\hline
10 & 9 & 1  &6 \\
\hline
10 & 10 & 1  &6\\
\hline
\hline
12 & 6 & 22 & 23$^{\star}$\\
\hline
\end{tabular}
\end{tiny}
\caption{Distance 6.
Note: The value superscripted with ``$\star$'' was
obtained through greedy search.\label{tab:dist6}}
\end{center}
\end{table}

\begin{table}[!t]
\begin{center}
\begin{tiny}
\begin{tabular}{|c|c||c||c|}
\hline
Length & Weight& Constant & Bounded\\
\hline
8 & 5  & 1 & 2\\
\hline
8 & 6  & 1  & 2\\
\hline
8 & 7  & 1 & 2\\
\hline
8 & 8  & 1  & 2\\   
\hline
\hline
9 & 5  & 2 & 2\\
\hline
9 & 6  & 1  &2\\
\hline
9 & 7  & 1  & 2\\
\hline
9 & 8  & 1  & 2\\
\hline
9 & 9  & 1  & 2\\
\hline
\hline
10 & 5 & 2 & 2\\
\hline
10 & 6 & 2  & 2\\
\hline
10 & 7 & 1  &2\\
\hline
10 & 8 & 1  &2\\
\hline
10 & 9 & 1  &2\\
\hline
10 & 10 & 1  &2\\
\hline
\hline
11 & 5 & 2  &2\\
\hline 
11 & 6 & 2  &2\\
\hline
11 & 7 & 2  &2\\
\hline
11 & 8 & 1  &2\\
\hline 
11 & 9 & 1  &2\\
\hline
11 & 10 & 1  & 2\\
\hline
11 & 11 & 1  & 2\\
\hline
\hline
12 & 5 & 3&3\\
\hline
\hline
13 & 5  & 3  & 3\\
\hline
\hline
14 & 7 & 8 & 8$^{\star}$ \\
\hline
\end{tabular}
\end{tiny}
\caption{Distance 8.
Note: The value superscripted with ``$\star$'' was
obtained through greedy search.\label{tab:dist8}}
\end{center}
\end{table}

We now discuss more broadly our results.  As noted above, 
in most cases the best bounded-weight codes we obtain are in
fact such that 
codes of optimal sizes are also provided by ``patching together''
existing optimal constant-weight codes.  However, this does not mean
that that part of our paper makes no contribution.  
Before our paper, it remained possible that there existed
bounded-weight codes for these cases having size larger than the
patched-together codes.  Our paper, via in many cases 
(namely, in all table lines other than the nine superscripted 
with asterisks) establishing the maximum size achievable 
by {\em any\/} legal code obeying parameters, removes this possibility.
Additionally, our work shows that in some cases the obvious patching
together that we mention does not achieve a maximum-sized code.  For
example, the case 
$A(4,8,6)$ is of this sort.  

We now turn to the question of whether, in light of our results,
bounded-weight codes seem wise to use.  Bounded-weight codes obviously
give no fewer codewords (in a maximum-sized code) that their sister
constant-weight codes.  Our tables show that in many cases they give
strictly more words.  Of course, as $w$ increases beyond 
$\lfloor
n/2\rfloor$ the size of the word-space of bounded-weight codes becomes
extremely rich relative to that of constant-weight codes (which starting at
weight 
$\lceil
n/2\rceil$
have contracting word-spaces as $w$ increases), and even for smaller
(but nonzero) values of $w$ their word space is of course richer---which
is exactly what opens up the possibility of larger-sized codes.

However, this does not necessarily mean that it is wise to use
bounded-weight codes.  As our results show, even maximum-sized
bounded-weight codes give scant improvement over their sister
constant-weight codes, at least in the range---$w \leq \lfloor
n/2\rfloor$---in which the bounded-weight codes don't have a
prohibitively unfair advantage in search-space size.  Indeed, in this
range, the increase in code size we found is disappointing, and 
as our codes in this range are all maximum-sized, this
disappointment reflects the actual, optimal state of such codes.
Additionally, there is a huge cost in adopting bounded-weight codes.
In particular, the deepest direct advantage of constant-weight is
that their weight provides an extra type of error detection.
Bounded-weight codes sacrifice this extra line of protection.

However, as a final comment, we mention that maximum-sized codes may
have potential future uses in alternate models of
computation/communication.  Though this is currently hypothetical, it
is not entirely implausible.  Consider for example some future
alternate model of information (storage or) transmission---perhaps
biological, perhaps electrical, perhaps something else---in which each
(stored or) transmitted ``word'' has $n$ binary ``bits'' (which might
be represented via genetic material, or via charged particles in a
given location, or so on) but such that, due to constraints of the
(storage or) transmission medium, if more than $w$ of the bits are
``on'' there is the possibility that the information in the word will
degrade, or that the computer or transmission
lines will incur physical damage.  Possible
reasons might include power limitations, heat dissipation, or
attraction between biological components.  In this admittedly
extremely hypothetical setting, bounded-weight codes might play a
valuable role, as their limitation would be exactly suited to the
physical constraints imposed by the (storage or) transmission medium.

\clearpage

\singlespacing

\appendix

\section{Appendix:  Codes} 

\newcommand{\codehead}[1]{\center{\small{#1}}}
\begin{multicols}{5}
\begin{tiny}

\begin{tabular}{c}
$A(6,4,3)$\\
\hline
000111 \\
011001 \\
101010 \\
110100 \\
\end{tabular}

\begin{tabular}{c}
$A(6,4,4)$\\
\hline
000000\\
010111\\
101011\\
111100\\
\end{tabular}

\begin{tabular}{c}
$A(6,4,5)$\\
\hline
000000\\
011110\\
111001\\
100111\\
\end{tabular}

\begin{tabular}{c}
$A(6,4,6)$\\
\hline
000000\\
001111\\
110011\\
111100\\
\end{tabular}

\begin{tabular}{c}
$A(7,4,3)$\\
\hline
0000111\\
0011001\\
0101010\\
0110100\\
1001100\\
1010010\\
1100001\\
\end{tabular}

\begin{tabular}{c}
$A(7,4,4)$\\
\hline
0000000\\
0101011\\
0110101\\
1011001\\
1101100\\
1110010\\
1000111\\
0011110\\
\end{tabular}

\begin{tabular}{c}
$A(7,4,5)$\\
\hline
0000000\\
0101101\\
1010101\\
0110011\\
0011110\\
1001011\\
1100110\\
1111000\\
\end{tabular}

\begin{tabular}{c}
$A(7,4,6)$\\
\hline
0000000\\
0001111\\
0110011\\
0111100\\
1010101\\
1011010\\
1100110\\
1101001\\
\end{tabular}

\begin{tabular}{c}
$A(7,4,7)$\\
\hline
0000000\\
0001111\\
0110011\\
0111100\\
1010101\\
1011010\\
1100110\\
1101001\\
\end{tabular}

\begin{tabular}{c}
$A(8,4,3)$\\
\hline
00000001\\
00101010\\
00110100\\
01001100\\
01010010\\
10011000\\
10000110\\
11100000\\
\end{tabular}

\begin{tabular}{c}
$A(8,4,4)$\\
\hline
00000000\\
00111100\\
01011010\\
01101001\\
10010110\\
10100101\\
11000011\\
00110011\\
01010101\\
01100110\\
10011001\\
10101010\\
11001100\\
00001111\\
11110000\\
\end{tabular}

\begin{tabular}{c}
$A(8,4,5)$\\
\hline
00000000\\
00111100\\
11011000\\
11100100\\
01001101\\
01010110\\
01101010\\
01110001\\
10001110\\
10010101\\
10101001\\
10110010\\
00011011\\
00100111\\
11000011\\
\end{tabular}

\begin{tabular}{c}
$A(8,4,6)$\\
\hline
00001111\\
00110011\\
01010101\\
01101001\\
10010110\\
10101010\\
11001100\\
11110000\\
00011000\\
00100100\\
01000010\\
01111110\\
10000001\\
10111101\\
11011011\\
11100111\\
\end{tabular}

\begin{tabular}{c}
$A(8,4,7)$\\
\hline
00001111\\
00110011\\
01010101\\
01101001\\
10010110\\
10101010\\
11001100\\
11110000\\
00011000\\
00100100\\
01000010\\
01111110\\
10000001\\
10111101\\
11011011\\
11100111\\
\end{tabular}

\begin{tabular}{c}
$A(8,4,8)$\\
\hline
00001111\\
00110011\\
01010101\\
01101001\\
10010110\\
10101010\\
11001100\\
11110000\\
00011000\\
00100100\\
01000010\\
01111110\\
10000001\\
10111101\\
11011011\\
11100111\\
\end{tabular}

\begin{tabular}{c}
$A(9,4,3)$\\
\hline
000000111\\
000011001\\
000101010\\
001001100\\
100010100\\
100100001\\
101000010\\
011000001\\
010100100\\
110001000\\
001110000\\
010010010\\
\end{tabular}

\begin{tabular}{c}
$A(9,4,4)$\\
\hline
000000000\\
001101010\\
010101100\\
011000110\\
100111000\\
000001111\\
000110011\\
101001100\\
101000011\\
001011001\\
111010000\\
110100010\\
010011010\\
110001001\\
011100001\\
100010110\\
001110100\\
100100101\\
010010101\\
\end{tabular}

\begin{tabular}{c}
$A(9,4,5)$\\
\hline
000000000\\
001101010\\
010101100\\
011000110\\
100111000\\
000001111\\
000110011\\
101001100\\
101000011\\
001011001\\
111010000\\
110100010\\
010011010\\
110001001\\
011100001\\
100010110\\
001110100\\
100100101\\
010010101\\
\end{tabular}

\begin{tabular}{c}
$A(9,4,6)$\\
\hline
000000000\\
001101010\\
010101100\\
011000110\\
100111000\\
000001111\\
000110011\\
101001100\\
101000011\\
001011001\\
111010000\\
110100010\\
010011010\\
110001001\\
011100001\\
100010110\\
001110100\\
100100101\\
010010101\\
\end{tabular}

\begin{tabular}{c}
$A(9,4,7)$\\
\hline
000000000\\
001101010\\
010101100\\
011000110\\
100111000\\
000001111\\
000110011\\
101001100\\
101000011\\
001011001\\
111010000\\
110100010\\
010011010\\
110001001\\
011100001\\
100010110\\
001110100\\
100100101\\
010010101\\
\end{tabular}

\begin{tabular}{c}
$A(9,4,8)$\\
\hline
000000000\\
001101010\\
010101100\\
011000110\\
100111000\\
000001111\\
000110011\\
101001100\\
101000011\\
001011001\\
111010000\\
110100010\\
010011010\\
110001001\\
011100001\\
100010110\\
001110100\\
100100101\\
010010101\\
111011111\\
\end{tabular}

\begin{tabular}{c}
$A(9,4,9)$\\
\hline
000000000\\
001101010\\
010101100\\
011000110\\
100111000\\
000001111\\
000110011\\
101001100\\
101000011\\
001011001\\
111010000\\
110100010\\
010011010\\
110001001\\
011100001\\
100010110\\
001110100\\
100100101\\
010010101\\
111111011\\
\end{tabular}

\begin{tabular}{c}
$A(10,4,3)$\\
\hline
0000000001\\
0000101010\\
0000110100\\
0001001100\\
1100000100\\
1001100000\\
0101000010\\
0010000110\\
0011010000\\
0110100000\\
1010001000\\
0100011000\\
1000010010\\
\end{tabular}

\begin{tabular}{c}
$A(10,4,4)$\\
\hline
0000100111\\
0010110001\\
0010101010\\
0000011110\\
0011000011\\
0001011001\\
0001101100\\
0001110010\\
0010001101\\
1000110100\\
0110010010\\
1000010011\\
0100010101\\
1001000101\\
1010011000\\
0100111000\\
1000101001\\
1010000110\\
1011100000\\
0110100100\\
0100001011\\
0101100001\\
1001001010\\
0111001000\\
0101000110\\
0000000000\\
1110000001\\
1101010000\\
0011010100\\
1100100010\\
1100001100\\
\end{tabular}

\begin{tabular}{c}
$A(11,4,6)$\\
\hline
00110111010\\
00001111110\\
00101110011\\
10101111000\\
10111100010\\
10011011010\\
10000111011\\
00011101011\\
01011110010\\
01001111001\\
10011110001\\
00010111101\\
10010110110\\
00011010111\\
11001101010\\
10101001011\\
10001100111\\
01010011011\\
11010111000\\
11001010011\\
00111011001\\
10001011101\\
10110101001\\
11011001001\\
10110010011\\
10101010110\\
11010100011\\
10011101100\\
01101011010\\
10010001111\\
11000011110\\
10100101110\\
00100011111\\
10110011100\\
00111001110\\
11111010000\\
11011000110\\
11110001010\\
11100011001\\
11100110010\\
01011011100\\
10111000101\\
10100110101\\
01111000011\\
11010010101\\
01100101011\\
01001001111\\
01000110111\\
00110100111\\
11000101101\\
01010101110\\
11101100001\\
01100000000\\
00001100000\\
00000000011\\
10000001000\\
00010010000\\
11100000111\\
01100111100\\
01101010101\\
01111101000\\
01011100101\\
11001110100\\
00101101101\\
01110110001\\
01110001101\\
00111110100\\
01101100110\\
01110010110\\
11110100100\\
11101001100\\
\end{tabular}

\begin{tabular}{c}
$A(8,6,4)$\\
\hline
00000011\\
11110000\\
\end{tabular}

\begin{tabular}{c}
$A(8,6,5)$\\
\hline
00000001\\
01111100\\
\end{tabular}

\begin{tabular}{c}
$A(8,6,6)$\\
\hline
00000000\\
11111100\\
\end{tabular}

\begin{tabular}{c}
$A(8,6,7)$\\
\hline
00000000\\
01111110\\
\end{tabular}

\begin{tabular}{c}
$A(8,6,8)$\\
\hline
00000000\\
00111111\\
\end{tabular}

\begin{tabular}{c}
$A(9,6,4)$\\
\hline
000000011\\
110010100\\
001111000\\
\end{tabular}

\begin{tabular}{c}
$A(9,6,5)$\\
\hline
000000111\\
101110100\\
110011001\\
011101010\\
\end{tabular}

\begin{tabular}{c}
$A(9,6,6)$\\
\hline
000000000\\
111111000\\
001110111\\
110001111\\
\end{tabular}

\begin{tabular}{c}
$A(9,6,7)$\\
\hline
000000000\\
111110001\\
011101110\\
100011111\\
\end{tabular}

\begin{tabular}{c}
$A(9,6,8)$\\
\hline
000000000\\
001111110\\
111001101\\
110110011\\
\end{tabular}

\begin{tabular}{c}
$A(9,6,9)$\\
\hline
000000000\\
000111111\\
111000111\\
111111000\\
\end{tabular}

\begin{tabular}{c}
$A(10,6,4)$\\
\hline
0000001111\\
0001110001\\
0110010010\\
1010100100\\
1101001000\\
\end{tabular}

\begin{tabular}{c}
$A(10,6,5)$\\
\hline
1111000001\\
0001011101\\
1000110011\\
0110011010\\
1100101100\\
0011100110\\
\end{tabular}

\begin{tabular}{c}
$A(10,6,6)$\\
\hline
0000000000\\
1111001100\\
0011010111\\
1100100111\\
1001111010\\
0110111001\\
\end{tabular}

\begin{tabular}{c}
$A(10,6,7)$\\
\hline
0000000000\\
0111101100\\
1010100111\\
1100111010\\
1011011001\\
0101010111\\
\end{tabular}

\begin{tabular}{c}
$A(10,6,8)$\\
\hline
0000000000\\
1101100011\\
1001111100\\
1110011010\\
0110101101\\
0011010111\\
\end{tabular}

\begin{tabular}{c}
$A(10,6,9)$\\
\hline
0000000000\\
0001111110\\
1110001110\\
0111010101\\
1011101001\\
1100110011\\
\end{tabular}

\begin{tabular}{c}
$A(10,6,10)$\\
\hline
0000000000\\
0000111111\\
0111000111\\
1110110001\\
1101101010\\
1011011100\\
\end{tabular}

\begin{tabular}{c}
$A(12,6,6)$\\
\hline
010010111100\\
011100101001\\
001000011111\\
011001110010\\
011110000110\\
001011100101\\
110000100111\\
111010010001\\
000110110011\\
100011010110\\
000000000000\\
101100110100\\
000101101110\\
010011001011\\
100001111001\\
010101010101\\
110111100000\\
100110001101\\
110100011010\\
111001001100\\
001111011000\\
101101000011\\
101010101010\\
\end{tabular}

\begin{tabular}{c}
$A(8,8,5)$\\
\hline
00000111\\
11111000\\
\end{tabular}

\begin{tabular}{c}
$A(8,8,6)$\\
\hline
00000011\\
11111100\\
\end{tabular}

\begin{tabular}{c}
$A(8,8,7)$\\
\hline
00000101\\
11111010\\
\end{tabular}

\begin{tabular}{c}
$A(8,8,8)$\\
\hline
00100001\\
11011110\\
\end{tabular}

\begin{tabular}{c}
$A(9,8,5)$\\
\hline
000000111\\
111110000\\
\end{tabular}

\begin{tabular}{c}
$A(9,8,6)$\\
\hline
000000111\\
111110000\\
\end{tabular}

\begin{tabular}{c}
$A(9,8,7)$\\
\hline
000000001\\
111111100\\
\end{tabular}

\begin{tabular}{c}
$A(9,8,8)$\\
\hline
000000000\\
111111110\\
\end{tabular}

\begin{tabular}{c}
$A(9,8,9)$\\
\hline
000000000\\
011111111\\
\end{tabular}

\begin{tabular}{c}
$A(10,8,5)$\\
\hline
1111100000\\
0000000111\\
\end{tabular}

\begin{tabular}{c}
$A(10,8,6)$\\
\hline
0011111110\\
1100010000\\
\end{tabular}

\begin{tabular}{c}
$A(10,8,7)$\\
\hline
1111111000\\
0000010110\\
\end{tabular}

\begin{tabular}{c}
$A(10,8,8)$\\
\hline
0000000000\\
1111111100\\
\end{tabular}

\begin{tabular}{c}
$A(10,8,9)$\\
\hline
0000000000\\
0111111110\\
\end{tabular}

\begin{tabular}{c}
$A(10,8,10)$\\
\hline
1111100000\\
0000011111\\
\end{tabular}

\begin{tabular}{c}
$A(11,8,5)$\\
\hline
11111000000\\
00100101110\\
\end{tabular}

\begin{tabular}{c}
$A(11,8,6)$\\
\hline
00000000011\\
00011111100\\
\end{tabular}

\begin{tabular}{c}
$A(11,8,7)$\\
\hline
11111000000\\
00000100110\\
\end{tabular}

\begin{tabular}{c}
$A(11,8,8)$\\
\hline
00001111111\\
11110110001\\
\end{tabular}

\begin{tabular}{c}
$A(11,8,9)$\\
\hline
00001111111\\
11110110001\\
\end{tabular}

\begin{tabular}{c}
$A(11,8,10)$\\
\hline
00000000000\\
00111111110\\
\end{tabular}

\begin{tabular}{c}
$A(11,8,11)$\\
\hline
00000000000\\
00011111111\\
\end{tabular}

\begin{tabular}{c}
$A(12,8,5)$\\
\hline
000000000111\\
011100011000\\
100011110000\\
\end{tabular}

\begin{tabular}{c}
$A(13,8,5)$\\
\hline
0000000000111\\
0111000101000\\
1000111100000\\
\end{tabular}

\begin{tabular}{c}
$A(14,8,7)$\\
\hline
10010100111100\\
11100011101000\\
01101110010100\\
00110110100011\\
11001100001011\\
10011011010010\\
00111001001101\\
01000001110111\\
\end{tabular}

\end{tiny}
\end{multicols}


\begin{thebibliography}{EHSW87}

\bibitem[AJ]{dimacs:graph:solvers}
D.~Applegate and D.~Johnson.
\newblock Available from ftp://dimacs.rutgers.edu/pub/challenge/graph/solvers.

\bibitem[BHOS97]{brouwer:codes}
A.~Brouwer, K.~Hamalainen, P.~Ostergard, and N.~Sloane.
\newblock Bounds on mixed binary/ternary codes.
\newblock {\em IEEE Transactions on Information Theory}, IT-44(1):1334--1380,
  January 1997.

\bibitem[Bro]{binary:linear:codes}
A.~Brouwer.
\newblock Bounds on the minimum distance of linear codes.
\newblock http://www.win.tue.nl/math/dw/voorlincod.html.

\bibitem[BSSS90]{brouwer:table}
A.~Brouwer, J.~Shearer, N.~Sloane, and W.~Smith.
\newblock A new table of constant weight codes.
\newblock {\em IEEE Transactions on Information Theory}, IT-36(6):1334--1380,
  November 1990.

\bibitem[CP90]{car-par:j:exact-cliques}
R.~Carraghan and P.~Paradalos.
\newblock An exact algorithm for the maximum clique problem\typeout{MINOR
  PANIC: car-par: Missing issue number}.
\newblock {\em Operations Research Letters}, 9:375--382, 1990.

\bibitem[CPS92]{binary:self:dual:codes}
J.~Conway, V.~Pless, and N.~Sloane.
\newblock The binary self-dual codes of length up to 32: A revised enumeration.
\newblock {\em J. Combinatorial Theory, Series A}, 60(2):183--195, 1992.

\bibitem[EHSW87]{gam-hem-shp-wei:j:annealing}
A.~{El Gamel}, L.~Hemachandra, I.~Shperling, and V.~Wei.
\newblock Using simulated annealing to design good codes.
\newblock {\em IEEE Transactions on Information Theory}, IT-33(1):116--123,
  1987.

\bibitem[JAMS91]{ara-joh-mcg-sch:j:cliques-etc}
D.~Johnson, C.~Aragon, L.~McGeoch, and C.~Schevon.
\newblock Optimization by simulated annealing: An experimental
  evaluation---{Part} {II}, graph coloring and number partitioning.
\newblock {\em Operations Research}, 39(3):378--406, 1991.

\bibitem[JT93]{dimacs:clique}
D.~Johnson and M.~Trick, editors.
\newblock {\em Cliques, Coloring and Satisfiability: {S}econd {DIMACS}
  Implementation Challenge}, number~26 in DIMACS series in Discrete Mathematics
  and Theoretical Computer Science. A.M.S., 1993.

\bibitem[LRS]{binary:codes:table}
S.~Litsyn, E.~Rains, and N.~Sloane.
\newblock Table of nonlinear binary codes.
\newblock Available at http://www.research.att.com/$\tilde{~}$njas/codes/And.

\bibitem[RS]{constant:weight:codes:table}
E.~Rains and N.~Sloane.
\newblock Table of constant weight binary codes.
\newblock Available at http://www.research.att.com/$\tilde{~}$njas/codes/Andw/.

\bibitem[Slo]{spherical:codes:tables}
N.~Sloane.
\newblock Available from http://www.research.att.com/$\tilde{~}$njas/.

\bibitem[VAvL93]{vaessens:gen-alg}
R.~Vaessens, E.~Aarts, and J.~van Lint.
\newblock Genetic algorithms in coding theory: {A} table for ${A}_3(n,d)$.
\newblock {\em Discrete Applied Mathematics}, 45(1):71--87, August 1993.

\end{thebibliography}
\end{document}